\documentclass[aip,reprint]{revtex4-1}
\usepackage{graphicx}

\begin{document}

\title{Characterization of individual layers in a bilayer electron system produced in a wide quantum well}

\author{S.\,I.~Dorozhkin}
\email[]{dorozh@issp.ac.ru}
\author{A.\,A.~Kapustin}
\author{I.\,B.~Fedorov}
\affiliation{Institute of Solid State Physics RAS, 142432
Chernogolovka, Moscow district, Russia}
\author{V.~Umansky}
\affiliation{Department of Physics, Weizmann Institute of Science,
76100 Rehovot, Israel}
\author{K.~von~Klitzing}
\author{J.\,H.~Smet}
\affiliation{Max-Planck-Institut f\"ur Festk\"orperforschung,
Heisenbergstrasse 1, D-70569 Stuttgart, Germany}

\date{\today}

\begin{abstract}
Here we report on a transparent method to characterize individual
layers in a double-layer electron system which forms in a wide
quantum well and to determine their electron densities. The
technique relies on the simultaneous measurement of the
capacitances between the electron system and gates located on
either side of the well. Modifications to the electron wave
function due to the population of the second subband and
appearance of an additional electron layer can be detected. The
magnetic field dependence of these capacitances is dominated by
quantum corrections caused by the occupation of Landau levels in
the nearest electron layer. The technique should be equally
applicable to other implementations of a double layer electron
system.
\end{abstract}

\maketitle

\section{INTRODUCTION}

Double-layer electron systems (DLESs) composed of a pair of
coupled two-dimensional electron systems (2DESs) have attracted
considerable attention due to the rich variety of collective
phenomena they host~\cite{Ref1}. Of particular interest has been
the equilibrium superfluid exciton condensate which emerges when
the total number of electrons in the double layer system equals
the degeneracy of a single spin split Landau level (see review
article~\cite{Eisen2} and Refs. therein). While nearly all
previous experiments were performed on GaAs/AlGaAs
heterostructures, this activity has recently received renewed
impulse from both theoretical~\cite{MacDon,Perali} and
experimental~\cite{Tutuc,Kim} studies on graphene-based devices.

In GaAs/AlGaAs heterostructures the double layer system is formed
either in a double quantum well device~\cite{Boebinger} or in a
wide quantum well (QW) exhibiting a confinement potential with two
minima located near the left and right barrier~\cite{Suen}. In the
wide quantum well implementation, the electron layers can not be
contacted separately, since annealed ohmic contacts short-circuit
the two layers. In double quantum well structures suitable methods
and sample designs have been developed~\cite{EisenMes} (see also
recent paper~\cite{Ritchie} and Refs. therein) to separately
contact the individual layers. This has enabled a far wider
variety of experiments including interlayer tunneling and electron
drag studies which have led to the discovery of a Bose-Einstein
exciton condensate in this double layer system.

Because the fabrication of separately contacted double layer
systems remains challenging, many studies continue to be carried
out using simultaneously contacted 2DESs. In such cases, an
essential task is the determination of the electron densities in
the individual layers or subbands. This is usually attempted by
analyzing the Shubnikov-de Haas oscillations (SdHO), which are
periodic in the inverse of the magnetic field, using Fourier
transform techniques. In a single 2DES the minima of SdHO
correspond to filling of integer numbers $i$ of spin split Landau
levels and their magnetic field positions $B_{\rm i}$ satisfy the
condition $n_{\rm s}=iLeB_{\rm i}/h$, where $n_{\rm s}$ is the
electron areal density, $eB/h$ represents the degeneracy of a
single spin split Landau level, and $L=1$ or 2 depending on
whether spin splitting is resolved or not. The SdHO frequency $F$
in the inverse magnetic field is equal to $F=hn_{\rm s}/eL$.
Therefore, in a single 2DES the fundamental frequency ($L=2$) is
observed at low magnetic fields when spin splitting is not
resolved, whereas at higher field strength the harmonic with $L=1$
may appear. In DLESs the number of Fourier harmonics increases and
the frequencies determined by the electron densities of the
individual layers are complemented by their sum and difference.
Namely, the frequency $F_{\rm diff}\propto |n_{\rm s1}-n_{\rm
s2}|$ (here $n_{\rm s1}$ and $n_{\rm s2}$ are electron densities
in two individual layers) originates from magnetointersubband
oscillations~\cite{Polyan,Coleridge,PRB46,Gusev,Raikh} brought
about by elastic intersubband scattering of electrons. The
frequency $F_{\rm sum}\propto n_{\rm s1}+n_{\rm s2}$ has been
associated with so-called single-layer behavior~\cite{Ritchie1996}
involving the redistribution of electrons among the
layers~\cite{Ritchie1996,Dorozh2016}. The relative strength of
different frequencies strongly depends on
temperature~\cite{PRB46,Gusev} since the magnetointersubband
oscillations in contrast to SdHO are not sensitive to the
temperature broadening of the Fermi distribution
function~\cite{Polyan,Raikh}. For a more detailed discussion of
the Fourier analysis of magnetoresistance oscillations we refer
the reader to the supplementary material~\cite{Supp} in
Ref.~\cite{Ensslin}.

\section{SAMPLES AND EXPERIMENTAL TECHNIQUE}

Capacitance measurements have been widely used for the
characterization of the distribution of electric charges in
semiconductors. In field-effect transistors containing a 2DES they
also allow studying the compressibility of the electronic
system~\cite{Stiles,Krav1,EisenPRL,Krav2,DorozhPRB2} as well as
the energy gaps of incompressible integer~\cite{Stiles,KhrpaiPRLI}
and fractional quantum Hall ground states that may
form~\cite{DorozhCap1993,Eisen1994,DorozhPRB1995,KhrpaiPRL}. The
capacitance technique was also previously used in experiments on
DLESs~\cite{EisenPRL,Ensslin2,Dolgopolov1,Dolgopolov2}, however
the compressibility of only one of the layers was measured. Here
we extend the capacitance method to get access to the
compressibility of the two individual layers in a wide quantum
well. We show that it is possible to detect the integer filling of
the Landau levels in each layer and, hence, to determine their
individual electron densities.

This study has been performed on a Hall bar sample processed from
a GaAs/AlGaAs heterostructure (see the experimental layout in
Fig.1) where the electron system resides in a 60 nm wide GaAs
quantum well (QW) located 140 nm below the sample surface. A
homogeneously doped in-situ grown GaAs layer 850 nm below the QW
served as a back gate. The QW was filled with electrons via
modulation doping of the top AlGaAs layer at a distance of 65 nm
from the QW. A Schottky front gate was created by evaporating a
thin gold film on the sample surface. By changing the dc voltages
between the electron system and the gates (front gate, $V_{\rm
FG}^{\rm dc}$, and back gate, $V_{\rm BG}^{\rm dc}$) the total
electron density $n_{\rm tot}$ was varied in the range $1.5\times
10^{11}\ {\rm cm}^{-2}\,-\,2.5\times 10^{11}\,{\rm cm}^{-2}$.
Increasing $V_{\rm BG}^{\rm dc}$ resulted in the population of the
second subband of the wide QW and gave rise to the formation of
the second layer closer to the back gate as schematically
illustrated in Fig.1. This will be confirmed experimentally below.
The ohmic contacts to the electronic system shaped into a Hall bar
enabled the acquisition of both the longitudinal and Hall
resistances (For the sake of simplicity, only one contact is shown
in Fig.1). The measurements were carried out with the sample
immersed in liquid $^3{\rm He}$ in the presence of a tunable
magnetic field perpendicular to the QW-plane.

\begin{figure}
\centering {\includegraphics{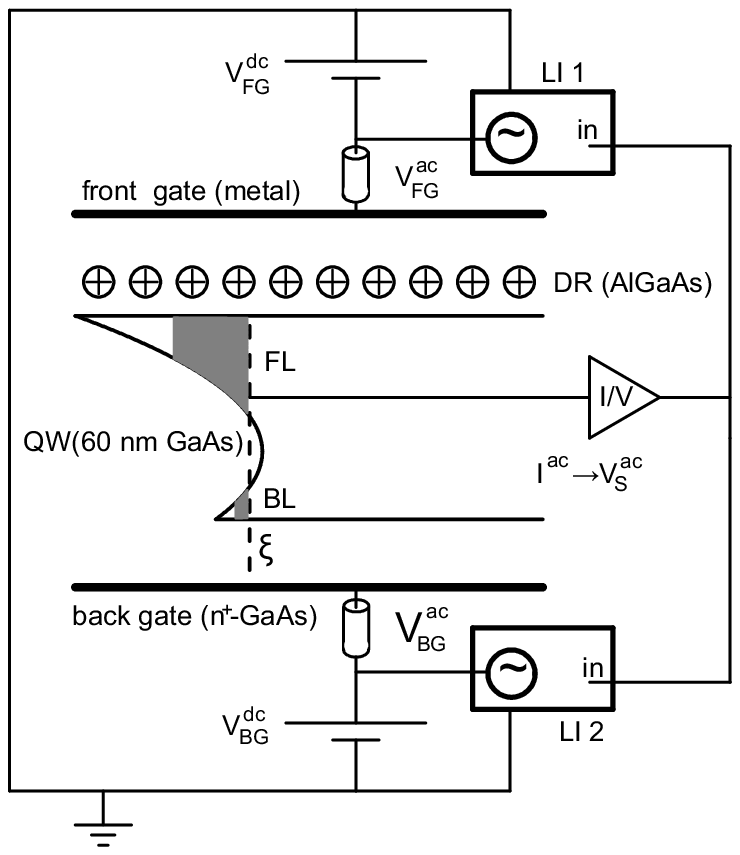}} \caption{Schematic of the
sample geometry and circuit layout for the capacitance
measurements. The occupied electron states in the two subbands
below the electrochemical potential $\xi$ are shown as dark shaded
regions. They form the front (FL) and back (BL) layer in the wide
GaAs quantum well (QW). DR denotes the modulation doped region.
LI1 and LI2 are dual-phase Lock-in amplifiers with internal
oscillators. The transimpedance amplifier converting the ac
current signal $I^{\rm ac}$ to an ac voltage $V^{\rm ac}_{\rm S}$
is shown as a right-pointing triangle.}
\end{figure}

\section{RESULTS AND DISCUSSION}

\begin{figure}
\centering {\includegraphics{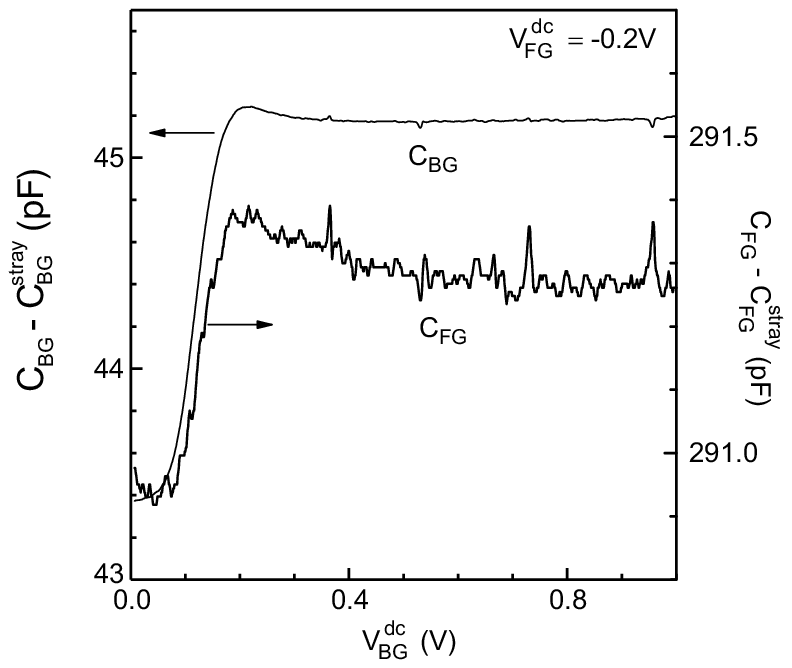}} \caption{The capacitances
measured between the electron system and back ($C_{\rm BG}$) and
front ($C_{\rm FG}$) gates as a function of the applied dc
back-gate voltage $V_{\rm BG}^{\rm dc}$. The stray capacitances
($C_{\rm BG}^{\rm stray}=27$~pF and $C_{\rm FG}^{\rm
stray}=17$~pF) have been subtracted from the measured values.
Measurements have been recorded at $T=0.5$~K.}
\end{figure}

The capacitances between the electronic system and both gates were
measured simultaneously using two dual-phase lock-in amplifiers
(LI1 and LI2). To measure capacitances, the gate voltages were
modulated with ac voltages $V_{\rm BG}^{\rm ac}=5\,{\rm mV}$ and
$V_{\rm FG}^{\rm ac}=1\,{\rm mV}$ with different frequencies
between 10 and 300 Hz produced by the internal oscillators of the
lock-in amplifiers. The induced ac currents were added and
converted to an ac voltage, $V_{\rm S}^{\rm ac}$, with a
transimpedance amplifier. The two frequency components were
detected separately with the two lock-in amplifiers. The
out-of-phase and in-phase signals from both amplifiers were
measured simultaneously. To minimize the stray capacitance, the
gates were connected to coaxial cables. However, some stray
capacitance remained in particular for the back gate which couples
not only to the electronic system but also to the ohmic contact
areas. This stray capacitance has been directly measured under
conditions where the integer quantum Hall effect is
well-developed. In this regime, the dissipative conductivity of
the 2DES tends to zero and the bulk of the electronic system is
not charged at the modulation frequencies used in the
experiment~\cite{DorozhJETPL1986}. For a highly conducting 2DES,
the out-of-phase signal is proportional to the capacitance. It
includes the quantum correction brought about by the finite
compressibility of the 2DES~\cite{Stiles} (for zero magnetic field
see also Ref.~\cite{Luryi}). For a field-effect transistor
composed of a 2DES in a narrow quantum well and a single gate the
equation for the capacitance $C$ reads as~\cite{Stiles},
\begin{equation}
\frac{S}{C}=\frac{S}{C_{\rm
g}}+\frac{1}{e^2}\frac{\partial\zeta}{\partial n_{\rm s}},
\label{cap}
\end{equation}
where $S$ is the gated area of the sample, $\zeta$ the chemical
potential of the 2DES and $C_{\rm g}=\epsilon_{\rm i} S/ d_0$ the
geometric capacitance with $d_0$ the thickness of the insulating
layer with dielectric permittivity $\epsilon_{\rm i}$ separating
the gate and the center of weight of the electron wave function in
the quantum well. The second term is inversely proportional to the
electronic compressibility $\kappa=(\partial n_{\rm s}/\partial
\zeta)/n^2_{\rm s}$. This term gives rise to capacitance minima
when the electronic system turns less compressible at magnetic
fields where the chemical potential is located in between Landau
levels, i.e. near integer fillings of the Landau levels.

The formation of a second layer or occupation of the second
subband is demonstrated in Fig.~2. Shown are variations of the
front and back gate capacitances with back gate voltage. Near
$V^{\rm dc}_{\rm BG}\approx 0.12$~V both capacitances display an
abrupt increase. The back (front) gate capacitance rises by about
1.8~pF (0.4~pF) or 4\% (0.14\%). In terms of a classical planar
capacitor this would correspond to a decrease of the distance
between the capacitor plates of approximately 34 nm (0.2 nm) for
the given distance of 850 nm (140 nm) between the QW and the back
(front) gate. These results imply that electrons start to occupy
the second subband and a new layer (back layer, BL) is formed,
located about 34 nm closer to the back gate than the ground
subband layer, i.e. front layer (FL)~\cite{Comment1}. At $B=0$,
the center of weight of the wave function in the ground subband is
hardly affected by the occupation of the second subband. It
practically does not change also when raising $V_{\rm BG}^{\rm
dc}>0.2$~V, i.e. with increasing electron density in the second
subband.

\begin{figure}
\centering {\includegraphics{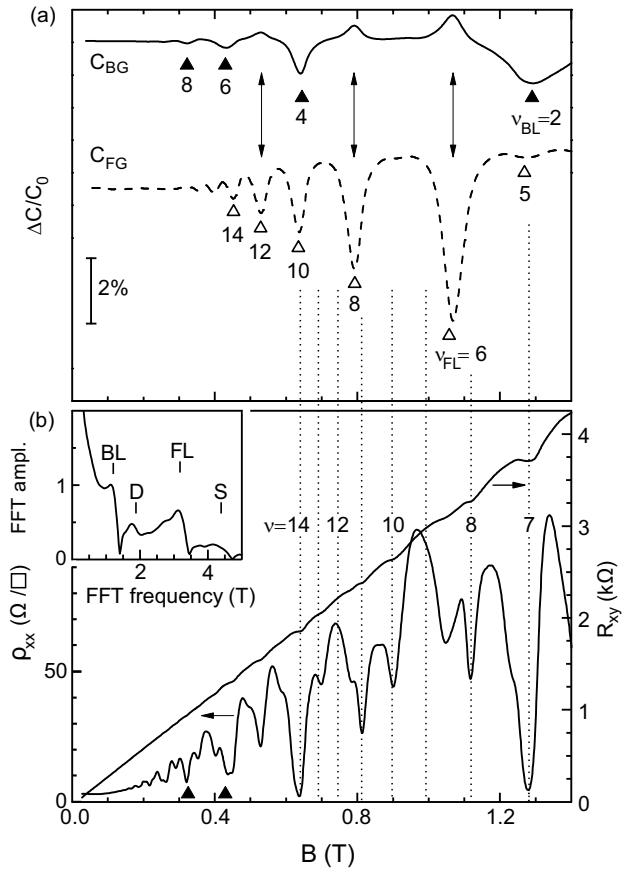}} \caption{(a) Variation of
the magnetocapacitances measured between the electron system and
the back ($C_{\rm BG}$, solid line) or front gate ($C_{\rm FG}$,
dashed line). The capacitances have been normalized by their value
$C_0$ at $B=0$ which has its own magnitude for each gate. The
closed (open) triangles mark the fields where an integer number of
levels in the back layer (front layer) is occupied and the number
indicates the filling factor $\nu_{\rm BL}$ ($\nu_{\rm FL}$) for a
2DES with an electron density $n_{\rm BL}=6.2\times 10^{10}\,{\rm
cm}^{-2}$ ($n_{\rm FL}=15.5\times 10^{10}\,{\rm cm}^{-2}$). The
triangles corresponding to $\nu_{\rm BL}=$\,6 and $\nu_{\rm
BL}=$\,8 are included in panel (b) as well. The double-sided
arrows mark the $C_{\rm FG}$ minima accompanied by maxima in
$C_{\rm BG}$. The magnetocapacitance data were measured at
T=1.5~K.(b) The magnetoresistance $\rho_{\rm xx}$ (left axis) and
Hall resistance $R_{\rm xy}$ (right axis) measured for the same
gate voltages $V^{\rm dc}_{\rm FG}=-0.2$~V and $V^{\rm dc}_{\rm
BG}=0.9$~V as the capacitance data in panel (a) and at T=0.5~K.
Dashed lines correspond to integer filling of a 2DES with a total
electron density $n_{\rm tot}=21.7\times 10^{10}\,{\rm cm}^{-2}$.
For some lines the filling factor has been included. The inset
displays a Fast Fourier Transform (FFT) of the magnetoresistance
oscillations. BL and FL mark the positions of the frequencies
corresponding to electron densities in the back and front layer,
respectively, as determined from the magnetocapacitance data in
panel (a). Letters S and D mark the sum and difference of these
frequencies.}
\end{figure}

Typical magnetocapacitance data are presented in Fig.~3(a). Each
curve shows one set of minima whose positions are periodic in the
inverse magnetic field. The periods are distinct for the $C_{\rm
BG}$ and $C_{\rm FG}$ curves. Based on the previous
magnetocapacitance studies of 2DES, we ascribe these
magnetocapacitance minima to the integer filling of the Landau
levels in the layer adjacent to a particular gate. From these
oscillations, we determine the areal electron densities in both
layers and mark the positions of the corresponding integer filling
factors using numbered triangles (see Fig.3(a)).

It is instructive to compare the magnetocapacitance data with the
magnetoresistance oscillations shown in Fig.3(b). Note the
complicated behavior of these oscillations at $B>0.6$~T. Some of
the magnetoresistance minima can be assigned to integer filling
($\nu=7,8,10,11,14$) of a 2DES with an electron density $n_{\rm
tot}=21.7\times 10^{10}\,{\rm cm}^{-2}=n_{\rm FL}+n_{\rm BL}$. The
same density $n_{\rm tot}$ is extracted from the Hall resistance.
However, there are no minima at $\nu=9,12,13$ and hence the
sequence of oscillations is not periodic in the inverse of the
magnetic field. For $B>0.6$~T the magnetoresistance minima in
general do not coincide with any of the magnetocapacitance
features. However, when the filling factor of both layers takes on
an integer value ($\nu_{\rm FL}=5$, $\nu_{\rm BL}=2$ and $\nu_{\rm
FL}=10$, $\nu_{\rm BL}=4$), the magnetoresistance shows deep
minima ($\nu=7$ and 14), which are accompanied by quantum Hall
plateaus. For $B<0.6$~T the oscillation pattern exhibits two
different frequencies. Two of the low frequency minima nearly
coincide with $\nu_{\rm BL}=6$ and $\nu_{\rm BL}=8$ as indicated
by the solid triangles in panel (b).

\begin{figure}
\centering {\includegraphics{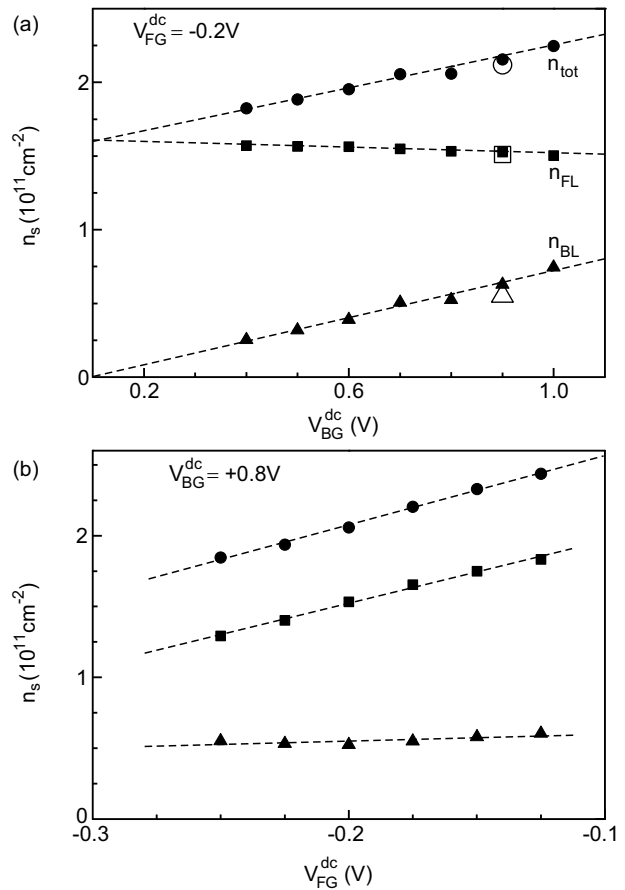}} \caption{ The areal
electron densities in the back ($n_{\rm BL}$, triangles) and front
($n_{\rm FL}$, squares) layers as a function of back-gate (a) and
front-gate (b) voltages as determined by the capacitance method.
The total electron density $n_{\rm tot}=n_{\rm BL}+n_{\rm FL}$ is
shown with circular symbols. The electron densities in the layers
determined from the Fourier spectrum in the inset to Fig.~3(b)
($V_{\rm FG}^{\rm dc}=-0.2$~V, $V_{\rm BG}^{\rm dc}=0.9$~V) are
displayed in Fig.4(a) as open symbols together with the total
electron density calculated from the Hall resistance of Fig.~3(b)
(open circle).}
\end{figure}

It is also informative to compare the electron densities obtained
from the magnetocapacitance data with those determined from the
Fourier spectrum analysis of the magnetoresistance oscillations
displayed in the inset to Fig. 3(b). Four maxima can be identified
in the spectrum. Two of them correspond to the oscillations from
the individual layers as they are located rather close to the
frequencies determined from the magnetocapacitance data  (marked
as BL and FL). Two other maxima lie at the sum (S) and difference
(D) of these frequencies. Hence, in general, the determination of
the electron densities from a Fourier spectrum requires careful
analysis. Our comparison of the two methods illustrates that the
magnetocapacitance data enable a straightforward interpretation
and ensure a markedly improved accuracy.

The electron densities determined from the magnetocapacitance data
are shown in Fig.4 for different front- and back-gate voltages.
The electron density in a layer increases linearly with the
voltage applied to the nearest gate and is only slightly affected
by the voltage on the gate separated by the other layer. In the
latter case the electron density dependence may even possess a
negative slope (see the $n_{\rm FL}(V_{\rm BG}^{\rm dc})$
dependence in Fig.4(a)). This is common in bilayer systems and has
been attributed~\cite{Millard,Smet} to exchange correlation
induced negative compressibility~\cite{EisenPRL} of 2D electron
systems at low density.

At first sight, the capacitance data look like those for two
independent 2DES. However, the effect of coupling between the
layers can be seen in Fig.3(a) as maxima on the $C_{\rm BG}$ curve
at the positions of some ($\nu_{\rm FL}=6,8,12$) minima of the
$C_{\rm FG}(B)$ dependence. These features are highlighted by
double-sided arrows.

\section{CONCLUSION}

In summary, we have shown that, compared to the magnetoresistance
measurements, our capacitance technique reveals effects of Landau
quantization in individual layers of a double-layer electron
system produced in a wide quantum well. In particular, this method
enables accurate determination of electron densities in each layer
despite the absence of separate contacts to each layer.

\begin{acknowledgments}
The experiment and data evaluation of this work were supported by
the Russian Foundation for Basic Research (Grant 17-02-00769). JHS
and VU acknowledge support from the GIF.
\end{acknowledgments}



\begin{references}
\bibitem{Comment1} We note that this estimate assumes infinite
electron compressibility in the layers, which affects its
accuracy.
\end{references}
\end{document}